\documentclass[journal]{IEEEtran}              

\usepackage{amsmath}
\usepackage{amssymb}
\usepackage{eqnarray}

\usepackage{graphicx}

\usepackage{cite}

\usepackage{subfig}
\usepackage{epsfig}

\usepackage{color}

\begin{document}
 \topmargin +9pt
\title{\huge{Computational neuroanatomy and gene expression:
  optimal sets of marker genes for brain regions}}

\author{Pascal Grange, Partha P. Mitra\\
Cold Spring Harbor Laboratory,\\
 One Bungtown Road, New York 11724, Cold Spring  Harbor, United States\\
Email: {\ttfamily{pascal.grange@polytechnique.org}}}
\maketitle


 \begin{abstract} 
The three-dimensional data-driven Anatomic Gene Expression Atlas
of the adult mouse brain consists of numerized {\emph{in situ}} hybridization 
data for thousands of genes, co-registered to the Allen Reference Atlas.
 We propose quantitative criteria to rank genes as markers of a brain region,
 based on the localization of the gene expression and on its functional fitting to the shape of the region. These criteria lead to natural generalizations
to sets of genes. We find sets of genes weighted with coefficients of both signs 
with almost perfect localization in all major regions of 
the left hemisphere of the brain, except the pallidum. Generalization of the fitting criterion with 
positivity constraint provides a lesser improvement of the markers, but requires sparser sets of genes.
 \end{abstract}

\begin{keywords} Gene expression, neuroanatomy, optimization, 
generalized eigenvalue problems.
 \end{keywords}

\section{Introduction: the Anatomic Gene Expression Atlas (AGEA) of the adult mouse brain}
 Neuroanatomy is experiencing 
a renaissance under the influence of molecular biology and computational methods.
 The Allen Institute has built a three-dimensional data-driven 
atlas of the adult mouse C57Bl/6J (see
see the NeuroBlast User Guide
{\ttfamily{http://mouse.brain-map.org/}}, and ~\cite{AllenGenome, AllenAtlasMol, neuroInfo, images}) containing expression data for thousands of genes,
co-registered to an atlas of brain regions,  the Allen Reference Atlas (ARA) \cite{AllenReferenceAtlas}.
However, there is no general agreement on the list of brain regions for rodents (see 
\cite{SwansonArchitecture,SwansonRat}).  Given an anatomical atlas such 
as the ARA, it is therefore natural to ask if brain regions can be 
recognised in the spatial patterns of gene-expression data. For a molecular approach to the anatomy of the hippocampus, see \cite{LeinHippocampus}.
 In the present note we propose quantitative criteria formalizing 
the notion of marker genes for brain regions.\\

\begin{figure}
\centering
\subfloat[]{\label{fig:Satb2Coronal}\includegraphics[width=2.7in]{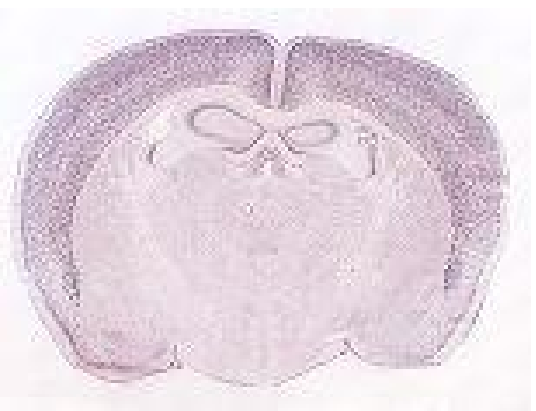}}\\
\subfloat[]{\label{fig:Satb2Proj}\includegraphics[width=2.7in]{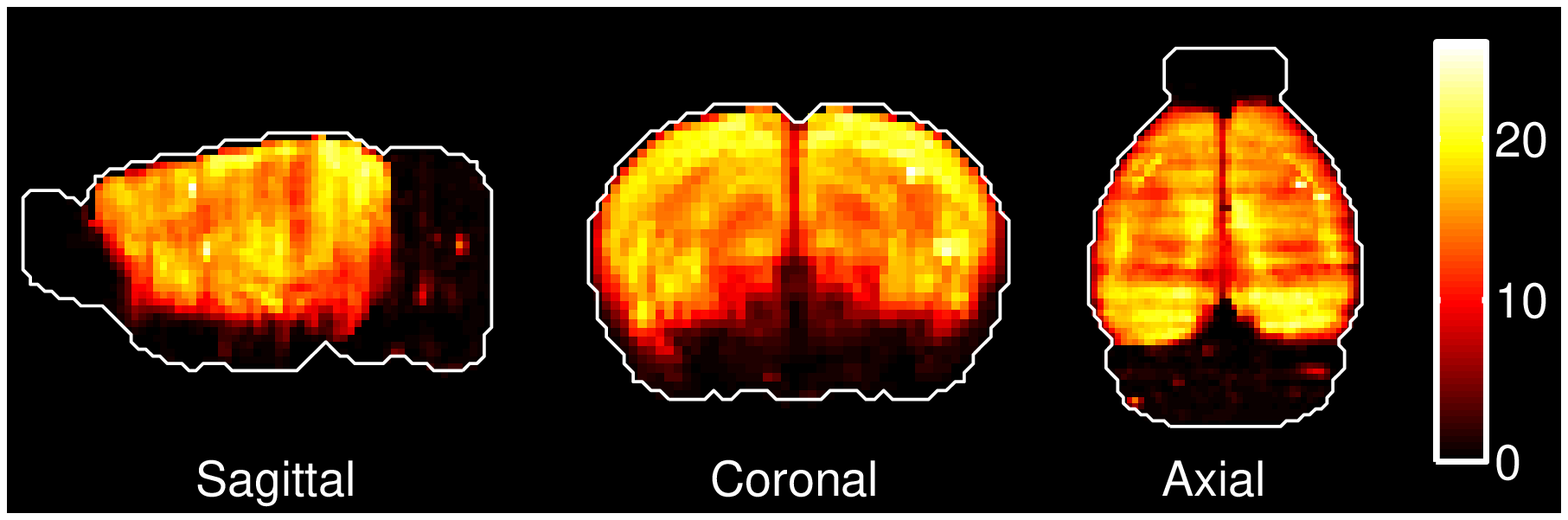}}\\
\caption{
{\bf{ISH-stained coronal slice of brain tissue and numerized data for {\emph{Satb2}}}}. (a) A coronal section of brain tissue. Colorimetric ISH gives rise to a blue precipitate where an mRNA for {\emph{Satb2}} is present. 
(b) A maximal-intensity projection of the three-dimensional data resulting from the co-registration 
of all coronal sections for {\emph{Satb2}} to a regular grid, at a resolution of 200 microns.}
\label{fig:Satb2}
\end{figure}

For each gene, an 
eight-week old C57Bl/6J male mouse brain was prepared as 
fresh-frozen tissue, and expression data were obtained through the following automated sequence 
of operations:\\
 1. Colorimetric {\it{in situ}}
    hybridization (a coronal section for {\emph{Satb2}} is shown on Figure \ref{fig:Satb2Coronal});\\
 2. Automatic processing of the resulting
    images: cell-shaped objects of size between 10 and 30 microns were looked for in each image in order 
to minimize artefacts;\\ 
3. Aggregation of the raw pixel data into a unique three-dimensional grid, with voxel side 200 microns (projections of the 
result is shown on Figure \ref{fig:Satb2Proj};\\
The  mouse brain is therefore partitioned into cubic voxels 
(the whole brain consists of $V=49,742$ voxels). For every voxel
$v$, the {\it{expression energy}} of the gene $g$ is defined as a
weighted sum of the greyscale-value intensities $I$ evaluated at the
pixels $p$ intersecting the voxel:
\begin{equation}
E(v,g) = \frac{\sum_{p\in v} M( p ) I(p)}{\sum_{p\in v} 1},
\label{ExpressionEnergy}
\end{equation}
 where
$M( p )$ is a Boolean mask worked out at step 2 with value 1 if the gene is expressed at pixel $p$ and 0 if it is not. A maximal-intensity 
projection of the gene-expression energy of {\emph{Satb2}} is shown on Figure \ref{fig:Satb2Proj}. The expression energy $E(v,g)$ is therefore 
expected to be proportional to the quantity of mRNA of gene $g$ in voxel $v$ (there can be saturation of the expression energy at large
values, but the expression energy is still a monotonic function of the total number of molecules of mRNA in the voxel).\\
For numerical applications we focused on a set of genes for which sagittal and coronal 
 sections have been produced at the Allen Institute. For each of these genes,
we computed the correlation between sagittal and coronal data. Some of these 
correlations are negative, and we chose to focus on three quarters of the genes ($G=3,041$ genes),
that make up the top of the distribution of correlation coefficients.The gene-expression data
we consider therefore consist of a voxel by gene matrix $E$ defined in Equation \ref{ExpressionEnergy}.
 Moreover,  the Allen Reference Atlas is registered to the same grid as the 
gene-expression data, so that  each voxel in the brain is annotated according to which
region it belongs. The ARA comes with several partitions of the brain, of varying coarseness. In particular,
the left hemisphere is partitioned into 12 disjoint regions in the ARA (each of which has one connected component, see Table \ref{tableNumAbove} for a list 
of these regions, and Figures 3b and 4c for an illustration of the cerebral cortex and the midbrain, respectively).
 This partition is referred to as the {\ttfamily{Big12}}
annotation. In the present note we will focus on this annotation for definiteness.\\
For computational purposes, a brain region $\omega$ is therefore equivalent to a set of  row indices
in the matrix $E$, and to a (normalized) vector $\chi_\omega$ in the $V$-dimensional voxel space, where the row indices
 are the only non-zero entries:
\begin{equation}
\chi_\omega( v ) \Leftrightarrow v \in \omega, \;\;\;\;\;\;\sum_{v = 1}^V \chi_\omega(v )^2 = 1.
\label{characteristic}
\end{equation}
 Each gene corresponds to a column of the matrix $E$, 
which is also a vector in a $V$-dimensional space. A marker gene  is
therefore a gene for which  this vector is closely aligned with $\chi_\omega$. 
 In the next section we propose two quantitative criteria formalizing this notion.\\

\section{Neuroanatomy from gene expression: ranking genes as markers}
\subsection{Ranking genes by localization scores}
 Given a brain region $\omega$ of interest, let us define
 the localization score of a gene $g$ as the fraction of the (square of the) $L^2$
 norm of its expression energy that is contained in the region:
\begin{equation}
\lambda_\omega(g) = \frac{\sum_{v\in\omega} E(v,g)^2 }{ \sum_{v\in\Omega} E(v,g)^2 },
\label{localizationScore}
\end{equation}
where $\Omega$ denotes the whole brain. We chose the $L^2$-norm 
because it is easy to generalize to a linear combination of genes (see next section).\\   
 We computed the localization score of every gene in every region of the {\ttfamily{Big12}} annotation.
 These scores induce a ranking of
genes as markers of each brain region. A perfect marker of the region $\omega$ according to this criterion would have 
a score of 1. Going from a region to another region, one has to be careful when comparing 
 the values of the localization scores: as the volumes of the brain regions vary
across the atlas, the localization scores $\lambda_\omega$ are biased by the size of the region $\omega$. 
We need a reference in order to estimate how good a localization score 
is compared to what could be expected for a given brain region. For a fixed brain region $\omega$ we can use two references.\\
A gene is a better marker of $\omega$ than expected from a uniform expression if its score 
$\lambda_\omega(g)$ is larger than the uniform reference defined as
\begin{equation}
\lambda^{\mathrm{uniform}}_\omega = \frac{{\mathrm{Vol}}\;\omega}{{\mathrm{Vol}}\;\Omega}.
\end{equation}
  A data-driven reference is given by the localization score of the average gene-expression profile:
\begin{equation}
\lambda^{\mathrm{average}}_\omega = \frac{\sum_{v\in\omega} E^{\mathrm{average}}(v)^2}{\sum_{v\in\Omega} E^{\mathrm{average}}(v)^2 },
\end{equation}
\begin{equation}
E^{\mathrm{average}}(v) = \frac{1}{G}\sum_{g=1}^G E(v,g).
\end{equation}
A gene is a better marker of $\omega$ than expected from an average expression if its localization score 
$\lambda_\omega(g)$ is larger than $\lambda^{\mathrm{average}}_\omega$.\\
\begin{table}
\centering
\small{
\begin{tabular}{|p{3.9cm}|p{1.3cm}|p{1.3cm}|}
\hline
\textbf{Region $\omega$ (abbreviation in the Allen Reference Atlas)}&\textbf{Percentage of genes above $\lambda^{\mathrm{uniform}}_\omega$}&\textbf{Percentage of genes above $\lambda^{\mathrm{average}}_\omega$}\\\hline
\textbf{Cerebral cortex (COR)}&59&26\\\hline
\textbf{Olfactory areas (OLF)}&41&40\\\hline
\textbf{Hippocampal region (HIP)}&51&35\\\hline
\textbf{{\hbox{Retrohippocampal reg. (RHP)}}}&53&33\\\hline
\textbf{Striatum (STR)}&16&28\\\hline
\textbf{Pallidum (PAL)}&9&34\\\hline
\textbf{Thalamus (THA)}&20&38\\\hline
\textbf{Hypothalamus (HYP)}&15&33\\\hline
\textbf{Midbrain (MID)}&13&37\\\hline
\textbf{Pons (PON)}&20&43\\\hline
\textbf{Medulla (MED)}&30&47\\\hline
\textbf{Cerebellum (CER)}&22&40\\\hline
\end{tabular} 

}
\caption{Percentage of a set of 3,041 genes in the Anatomic Gene Expression Atlas above the uniform and average references for the 
 regions in the {\ttfamily{Big12}} annotation of the left hemisphere in the Allen Reference Atlas. There is no
particular solidarity between the two columns.}
\label{tableNumAbove}
\end{table}
\subsection{Ranking genes by fitting scores}
The localization score does not take into account the detailed repartition
of the expression energy inside the region of interest. It is therefore
 interesting to study another ranking of genes, that compares
the gene-expression profiles to characteristic functions of brain regions.
  Such a comparison can be based on the functional distance between the expression 
profile and the characteristic function of the region. Let us choose the $L^2$ distance and compute the following 
fitting score for each gene $g$ in a given region $\omega$:
\begin{equation}
\phi_\omega(g) = 1 - \frac{1}{2} \sum_{v\in \Omega}\left( E^{\mathrm{norm}}_g(v ) - \chi_\omega(v)\right )^2,
\end{equation}
where $E^{\mathrm{norm}}_g$ is the $L^2$-normalized $g$-th column $E_g$ of the matrix of gene-expression energies:
\begin{equation}
E^{\mathrm{norm}}_g( v ) = \frac{E(v,g)}{\sqrt{\sum_{v=1}^V E( v,g)^2}}.
\end{equation}
 It is also useful to consider $E_g$ as a vector in the $V$-dimensional voxel space (it is the gene-expression vector
of gene $g$).
Just as in the definition of localization scores, we could have chosen another norm,
but the $L^2$-norm yields an intersting geometric interpretation of the fitting score. 
Expanding the expression of the fitting score in powers of the gene-expression data yields the cosine of the angle
 between the gene-expression vector $E_g$ and the vector $\chi_\omega$  in voxel 
space. The fitting score is therefore very closely related to
the notion of co-expression (which for two genes can be defined as the cosine of the
angle between their expression vectors, which is a useful quantity to study 
in order to estimate collective properties of sets of genes \cite{coExpressionStats}).\\
A perfect
marker of the region $\omega$ would be a gene with fitting score equal
to 1.\\
There are conflicts between the two induced rankings of genes, for instance {\emph{Satb2}}
 has the highest fitting score for the cortex (and indeed by the look of Figure \ref{fig:Satb2} it
is a good marker of the cortex), whereas is is ranked 8 by the localization scores, with $\lambda_{\mathrm{cortex}}=0.9345$.
On the other hand, {\emph{Pak7}} is ranked first by localization score, and 7th by fitting score. See Figure \ref{fig:best} for 
a plot of best fitting and localization scores in the regions of the {\ttfamily{Big12}} annotation. Pallidum is the region for which the best
fitting and localization scores are the lowest, and cerebellum is the one for which there are the highest. 
\begin{figure}
\centering
\includegraphics[width=3in]{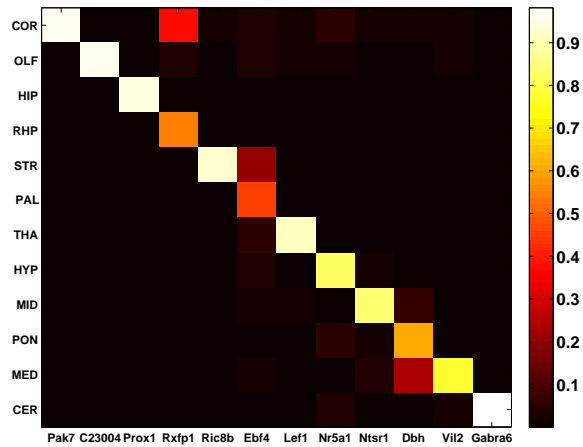}
\caption{{\bf{Localization scores of the best markers of each of the brain regions in the {\ttfamily{Big12}}
annotation.}} The $k$-th column contains the localization scores of the best marker of the $k$-th region, hence the diagonal look
of the figure. {\emph{Gabra6}}, the best marker of the  cerebellum, is the gene that maximizes the localization scores across all regions,
at 98.5 persent.}
\label{fig_bestLocalizationMatrix}
\end{figure}
\section{Sets of genes as markers}
\subsection{Generalized localization and generalized eigenvectors}
Looking at the scores of the top marker genes for each brain regions,
it appears that {\emph{Gabra6}} maximizes localization scores across all brain regions and all genes,
whereas the best marker in pallidum is the hardest to separate from other brain regions.
Hovever, comparing the numbers of genes localized above the average and uniform
reference values, as in Table \ref{tableNumAbove} does not show any particular ranking of
brain regions.\\
In order to find better markers, consider a linear superposition of expression energies in our dataset:
\begin{equation}
E_\alpha( v ) : = \sum_{g = 1}^{G} \alpha_g E( v, g ),
\label{eq:superposition}
\end{equation}
where $G=3,041$ is the number of genes in our dataset.\\
The localization score in the brain region $\omega$ of a weighted set
of genes  encoded by Equation \ref{eq:superposition} is naturally written as
\begin{equation}
         \lambda_\omega( \alpha ) = \frac{\sum_{v\in\omega} \left( \sum_g
  \alpha_g E( v, g )\right)^2 }{\sum_{v\in\Omega} \left( \sum_g \alpha_g
  E( v, g )\right)^2} = \frac{\alpha^t J^\omega \alpha}{\alpha^t  J^\Omega \alpha},
\label{lambdaOmega}
\end{equation}
 where the quadratic forms $J^\omega$ and
$J^\Omega$ have coefficients given respectively by scalar products of the projections of gene-expression vectors on $\omega$ and the whole brain:
 \begin{equation}
J^\omega_{g,h}= \sum_{v\in\omega} E(v,g)E(v,h),\; J^\Omega_{g,h}= \sum_{v\in_\Omega} E( v,g)E(v,h).
\end{equation}
The (generalized) localization score $\lambda_\omega( \alpha )$ is invariant under 
        multiplication of the vector $\alpha$. We can fix this dilation invariance by fixing the 
 value of the denominator in Equation \ref{lambdaOmega}. 
 Maximization the of 
        localization score boils therefore down to a maximization of the
        quadratic form $J^\omega$ under a quadratic constraint:
         \begin{equation}
 {\mathrm{max}}_{\alpha\in{\mathbf{R}}^G}\lambda_\omega(
         \alpha ) ={\mathrm{max}}_{\alpha\in{\mathbf{R}}^G, \alpha^t
           J^\Omega \alpha = 1}\alpha^t J^\omega\alpha.
\end{equation}
     Introducing
         the Lagrange multiplier $\sigma$ associated to the constraint,
         we are led to the maximization of the quadratic quantity
         \begin{equation}
        L_{\omega, \sigma}( \alpha ) = \alpha^t J^\omega\alpha - \sigma( \alpha^t J^\Omega \alpha - 1).
        \end{equation}
         The stationarity condition of $L_{\omega, \sigma}$ wrt the vector $\alpha$  yields a generalized eigenvalue problem,
       \begin{equation}
        J^\omega \alpha = \sigma J^\Omega \alpha,
       \label{eq:genEigenValue}
        \end{equation}
        and the maximum value of the generalized localization score is the largest generalized eigenvalue, while the associated 
        generalized eigenvectors contains the set of weights for genes in the best-localized superposition.\\
\begin{figure}
\centering
\subfloat[]{\label{fig:genProfile2}\includegraphics[width=2.7in]{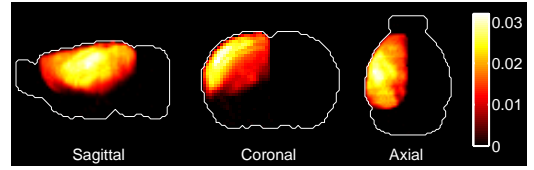}}\\
\subfloat[]{\label{fig:regionProfile2}\includegraphics[width=2.7in]{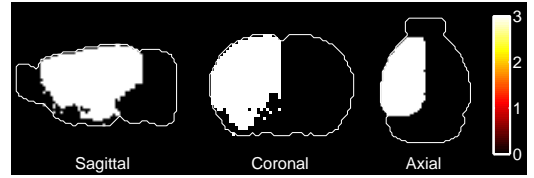}}\\
\subfloat[]{\label{fig:genVectors2}\includegraphics[width=2.7in]{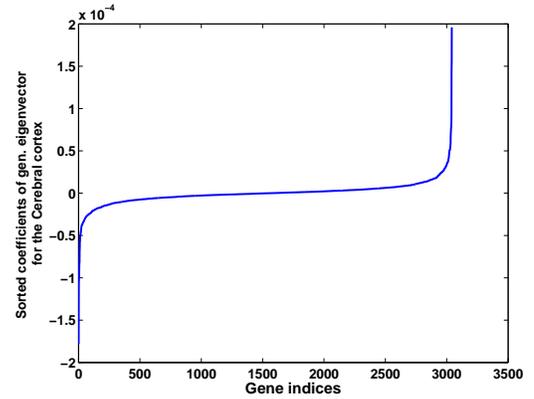}}\\
\caption{{\bf{The best set of genes as a generalized eigenvector for the cerebral cortex.}} (a) A maximal-intensity projection of the linear combination of the genes in the Adult Gene Expression Atlas corresponding to the generalized eigenvector that maximizes the localization in the cerebral cortex. 
(b) A maximal-intensity projection of the characteristic function of the cerebral cortex.  
(c) A plot of the sorted coefficients of the genes
in the generalized eigenvectors. The localization score in the cortex in 0.9994. {\emph{Pak7}} is at the second rank by its coefficient in the generalized eigenvector, while
{\emph{Satb2}} is only at the 64th rank.}
\end{figure}
The alternating signs of the coefficients make these sets
difficult to interpret in terms of transcriptional activity, and the plot of the sorted coefficients 
of the generalized eigenvector for the cerebral cortex in Figure 3c shows that the solutions are not sparse.  But these algebraic solutions provide absolute
bests that one could not beat by taking combinations of genes with
positive coefficients. The negative coefficients allow to offset the
contribution of some genes outside the region of interest.\\
\subsection{Generalized fitting scores and sets of genes weighted by positive coefficients}
Considering again a linear combination of gene-expression vectors, as in Equation \ref{eq:superposition}, but  weighted by  positive coefficients:
it is natural to propose 
the following fitting score, which just consists of the (square of) the $L^2$ distance 
between the normalized  sum of the expression energies of all genes in the set, and the characteristic function 
of the region of interest:
\begin{equation}
\phi_\omega(\alpha)  = 1 - \frac{1}{2} \sum_{v\in \Omega}\left( E^{\mathrm{norm}}_\alpha( v ) - \chi_\omega( v ) \right)^2,\; \alpha \in {\mathbf{R}}_+^G.
\label{genFitting}  
\end{equation}
A generalization of the fitting criterion to sets of genes that both solves
the sign problem and the sparsity problem is found quite 
naturally in terms of an $L^2$-$L^1$ minimization. The following function penalizes the 
$L^2$ error function of Equation (\ref{genFitting}) by the $L^1$ norm of the 
vector:
\begin{equation}
{\mathrm{ErrFit}}^{\omega,\Lambda}_{L^1-L^2}( \{\alpha\}) = || E^{\mathrm{norm}}_\alpha - \chi_\omega||^2_{L^2} + \Lambda||\alpha||_{L^1},
\label{eq:L1L2}
\end{equation}
which can be minimized wrt the weights of the genes \cite{L1L2} using Matlab code by K. Koh:
\begin{equation}
\alpha_\omega^\Lambda= {\mathrm{argmin}}_{\alpha \in {\mathbf{R}}_+^G}{\mathrm{ErrFit}}^{\omega,\Lambda}_{L^1-L^2}( \{\alpha\}).
\label{eq:optimPos}
\end{equation}
The range of parameter $\Lambda$ to be studied can be restricted to $[ 0, \Lambda_{\omega}^{\mathrm{max}} ]$, where 
\begin{equation}
 \Lambda_{\omega}^{\mathrm{max}} = 2 \;\mathrm{max}( E^t .\chi_\omega ),
\end{equation}
because for larger values of $\Lambda$, the quadratic form  ${\mathrm{ErrFit}}^{\omega,\Lambda}_{L^1-L^2}$ is bounded from below
by the squared norm of the vector $E_\alpha$, and the solution to the problem of Equation (\ref{eq:optimPos}) is trivially zero.
The best fitting score  is generally a decreasing function of $\Lambda$, while the sparsity grows with $\Lambda$. 
For each region $\omega$ in the {\ttfamily{Big12}} annotation, there is a domain of $[ 0,  \Lambda_{\omega}^{\mathrm{max}} ]$ for which the generalized fitting score $\phi_\omega( \alpha_\omega^\Lambda)$ is
larger than the best fitting score of a single gene (scores are plotted on Figure \ref{fig:best} for $\Lambda = 0.01  \Lambda_{\omega}^{\mathrm{max}}$).
\begin{figure}
\centering
\subfloat[]{\label{singleRegion10identifier5}\includegraphics[width=2.7in]{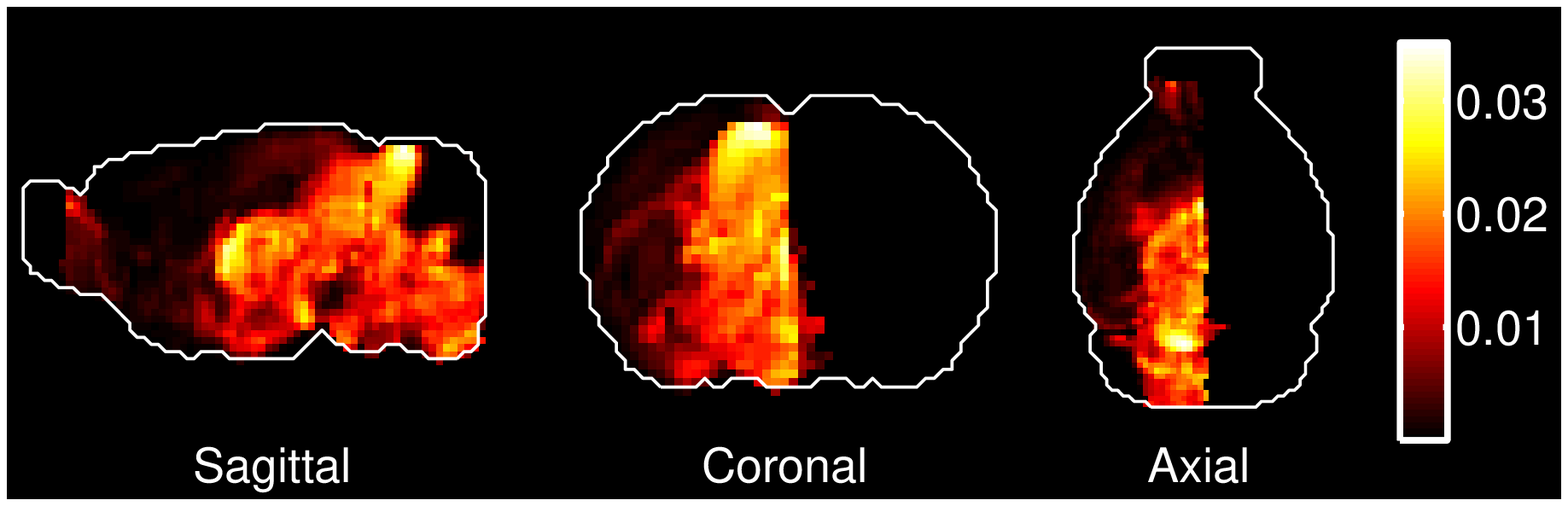}}\\
\subfloat[]{\label{setRegion10identifier5}\includegraphics[width=2.7in]{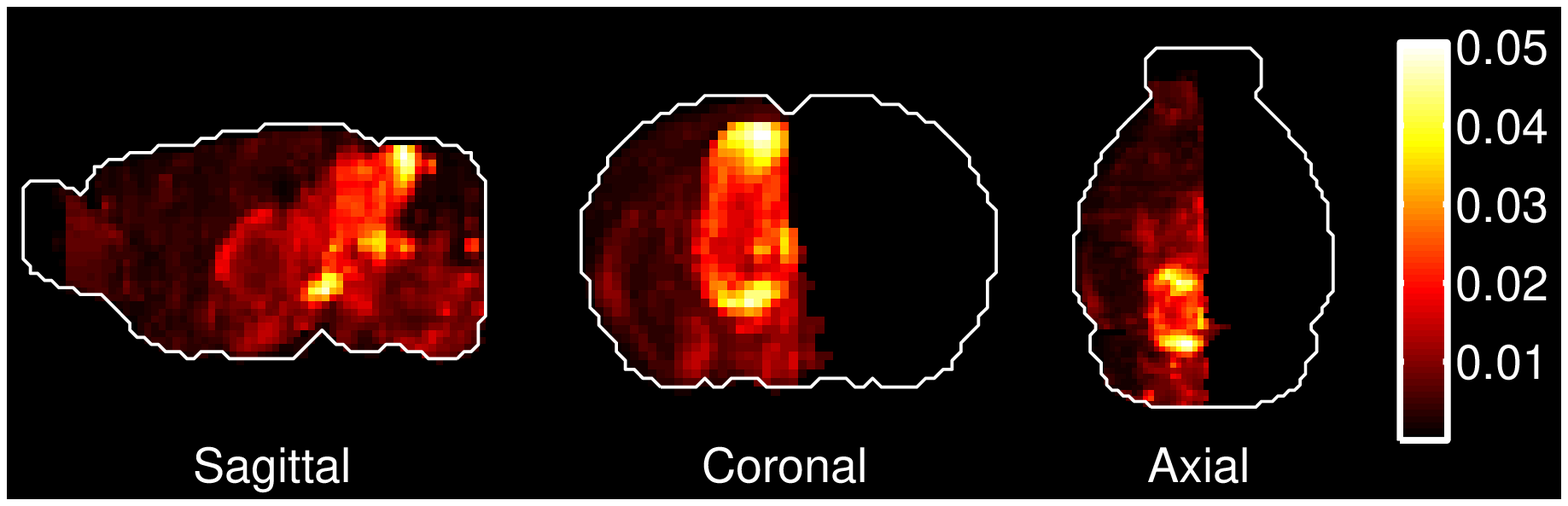}}\\
\subfloat[]{\label{profileRegion10identifier5}\includegraphics[width=2.7in]{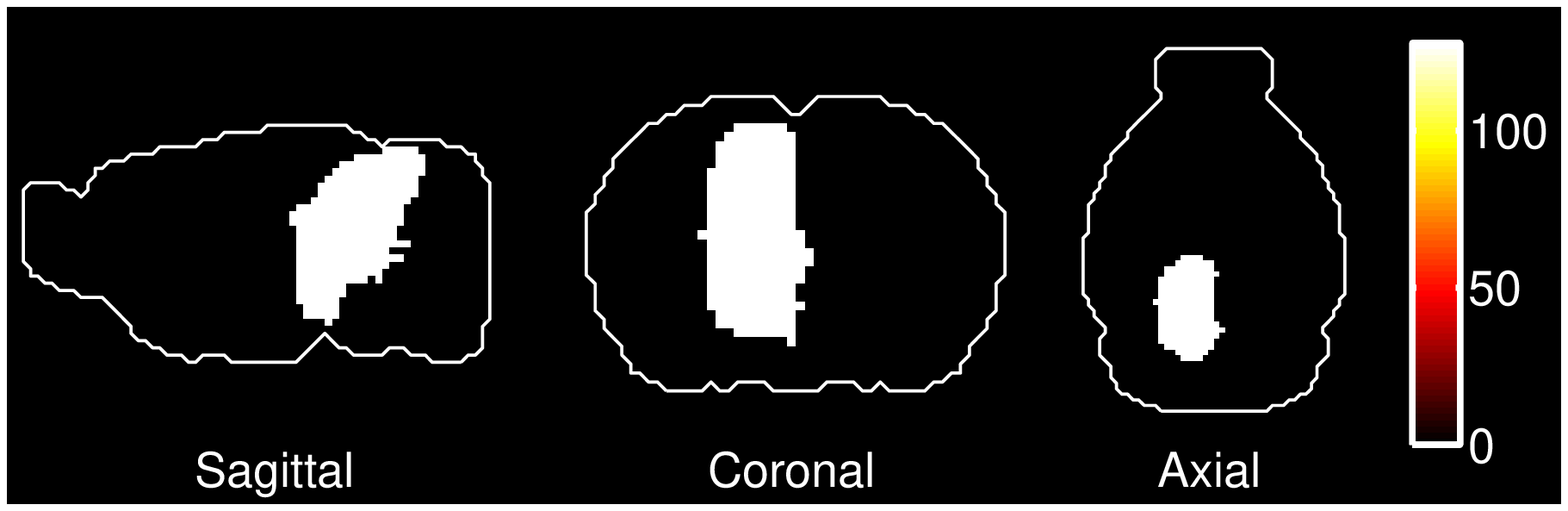}}\\
\caption{{\bf{Best markers by fitting for the midbrain.}} (a) The best single single gene is {\emph{Slc7a6}}; (b) the best set of genes consists of
 8 genes ({\emph{Slc17a6,    Ephb1,
    Sema3f,
    Glra3,
   Nova1,
    Tcf7l2,
    Ddc,
    Chrna6}}), at 
$\Lambda = 0.01 \times \Lambda_{\mathrm{midbrain}}^{\mathrm{max}}$; (c) Projection of
the midbrain.}
\label{fig:fittingsRegion10}
\end{figure}
\begin{figure}
\centering
\subfloat[]{\includegraphics[width=3.3in]{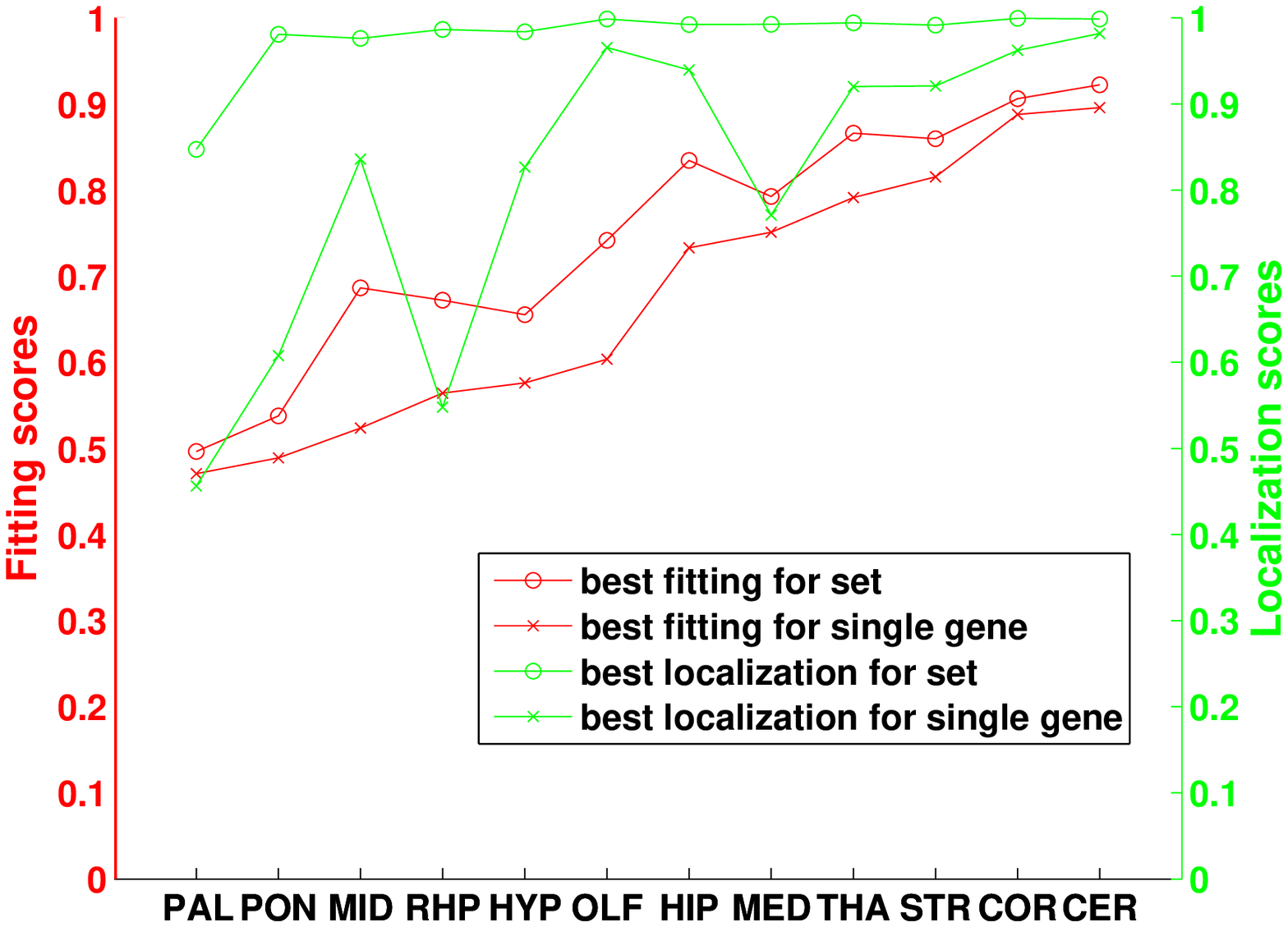}}\\
\subfloat[ ]{
\begin{tabular}{|p{1.5cm}|p{0.6cm}|p{0.6cm}|p{0.6cm}|p{0.6cm}|p{0.6cm}|p{0.6cm}|}
\hline
\textbf{\scriptsize{Brain region}}&PAL&PON&MID&RHP&HYP&OLF\\\hline
\textbf{\scriptsize{Best-fitting single genes}}&\scriptsize{{\emph{Ebf4}}}&\scriptsize{{\emph{Klk6}}}&\scriptsize{{\emph{Slc17a6}}}&\scriptsize{{\emph{Rspo2}}}&\scriptsize{{\emph{Gpr165}}}&\scriptsize{{\emph{Ppfibp1}}}\\\hline
\textbf{\scriptsize{Nb of genes in best-fitted set}}&7&8&8&9&9&12\\\hline
\end{tabular}
}\\
\subfloat[ ]{

\begin{tabular}{|p{1.5cm}|p{0.6cm}|p{0.6cm}|p{0.6cm}|p{0.6cm}|p{0.6cm}|p{1cm}|}
\hline
\textbf{\scriptsize{Brain region}}&HIP&MED&THA&STR&COR&CER\\\hline
\textbf{\scriptsize{Best-fitting single genes}}&{\emph{\tiny{TC1412430}}}&{\emph{\scriptsize{Glra1}}}&{\emph{\scriptsize{Lef1}}}&{\emph{\scriptsize{Rgs9}}}&{\emph{\scriptsize{Satb2}}}&{\emph{\tiny{3110001A13Rik}}}\\\hline
\textbf{\scriptsize{Nb of genes in best-fitted set}}&8&8&10&13&8&14\\\hline
\end{tabular}}\\
\caption{{{(a) Best fitting scores and localization scores of single genes and of best sets of genes for the brain regions
 of the {\ttfamily{Big12}} annotation.}} The brain regions are sorted by the best fitting score of single genes (see Table \ref{tableNumAbove} for the abbreviations
of the brain regions in the Allen Reference Atlas). The largest improvement to fitting brought by considering sets of genes rather 
than single genes in the midbrain. {{(b,c) Table of genes with highest fitting scores, and numbers of genes contributing to the best-fitted sets of genes.}}}
\label{fig:best}
\end{figure}
\section{Conclusions}
Quantitative methods used to rank single genes as markers 
of brain regions can efficiently spot genes whose expression profile
outlines a brain region of interest. In particular, the generalized localization score can yield almost 
perfectly localized gene expression for all the major  brain regions except pallidum, at the price of
involving thousands of genes, weighted by coefficients of both signs.
 The fitting criterion 
 can be generalized to sparse sets of genes with positive coefficients,
even though the improvement of the scores is less spectacular.\\
 The complexity of the taxonomy of cell types, and the precise anatomical localization 
in the brain of some of these cell types, indicates that there must be combinations of large numbers
of genes with positive coefficients, corresponding to the superposition 
of genes given by Equation \ref{eq:superposition}, that mark some brain regions, quite 
possibily much smaller than the large compartments of the left hemisphere we considered here \cite{foreBrainTaxonomy,RossnerCells,CahoyCells, DoyleCells,OkatyCells,SFN2011}.\\
The quantitative criteria used to define marker genes in the present paper are all global in
nature, since they all involve comparison of gene-expression vectors to brain regions
in terms of the entire voxel space. This does not
make use of the fact that the voxels belonging to the same region of the ARA form connected sets 
of the left hemisphere. One can make these methods more local  \cite{markerGenes} and
look for genes that are aligned to the projection of a brain region to a subspace of voxel space that surrounds the region. Such a set 
of voxels can be computed using level-sets of the eikonal distance to the boundary of the region \cite{LLMec,Sethian}. The eikonal distance
is also a useful geometric tool for the registration of mouse skulls to a reference skull, which is currently used in a 
high-throughput neuroanatomy project
 \cite{injections,MBA}. Genes separating brain regions from their environment without being particularly 
well localized or fitted globally are shown in \cite{markerGenes}. Moreover, the conservation properties of marker genes (or their lack of conservation properties)
 when going from the mouse atlas to  molecular atlases for other species will be relevant to the study of brain evolution\footnote{Co-expression coefficients
and localization scores for the human atlas can be found online at 
{\ttfamily{http://addiction.brainarchitecture.org}}}.\\

\section*{Acknowledgments}
It is our pleasure to thank Michael Hawrylycz for discussions. This research is supported by
 the NIH-NIDA Grant 1R21DA027644-01, {\emph{Co-expression networks of genes in the mouse and human brain}}.

 \end{document}